\documentclass[12pt,a4paper,pre,onecolumn,floatfix]{revtex4}
\usepackage[pdftex]{graphicx}
\usepackage{latexsym}
\usepackage{times}
\begin{document}
\title{One Hub-One Process: A Tool Based View on Regulatory Network Topology}

\author{Jacob Bock Axelsen}
\affiliation{{\small Centro de Astrobiolog\'{\i }a, Instituto Nacional de T\'ecnica
Aeroespacial, Ctra de Ajalvir km 4, 28850 Torrej\'on de Ardoz, Madrid,
Spain}}

\author{Sebastian Bernhardsson}
\affiliation{Department of Theoretical Physics, Ume{\aa} University,
901 87 Ume{\aa}, Sweden}

\author{Kim Sneppen\footnote{To whom correspondence should be
    addressed: sneppen@nbi.dk}} \affiliation{Center for Models of
  Life, Niels Bohr Institute, Blegdamsvej 17 DK-2100 Copenhagen {\O},
  Denmark}

\date{\today}

\begin{abstract}
  The relationship between the regulatory design and the functionality
  of molecular networks is a key issue in biology. Modules and motifs
  have been associated to various cellular processes, thereby
  providing anecdotal evidence for performance based localization on
  molecular networks. To quantify the structure-function relationship
  we investigate si\-mi\-la\-ri\-ties of proteins which are close in the
  regulatory network of the yeast Saccharomyces Cerevisiae. We find
  that the topology of the regulatory network show very weak remnants
  of its history of network reorganizations, but strong features of
  co-regulated proteins associated to similar tasks. This suggests
  that local topological features of regulatory networks, including
  broad degree distributions, emerge as an implicit result of matching
  a number of needed processes to a finite toolbox of proteins.
 \end{abstract}

\maketitle

\newpage

\section*{Introduction}
Contemporary systems biology have provided us with a large amount of
data on topology of molecular networks, thereby giving us glimpses
into computation and signaling in living cells. It have been found
that 1) regulatory networks have broad out-degree distributions
\cite{albert-review,maslov2005}, 2) transcriptional regulatory
networks contains many feed forward motifs \cite{milo}, and 3) highly
connected hubs are often found on the periphery of the network
\cite{maslov2002}. These findings are elements in understanding the
topology of existing molecular networks as the result of an interplay
between evolution and the processes they orchestrate in the cell.

In this paper we consider properties of proteins in the perspective of
how they are positioned relative to each other in the network.  This
is in part motivated by the existence of highly connected proteins
(hubs) and their relation to soft modularity
\cite{hartwell,maslov2002} in regulatory networks.  In particular one
may envision broad degree distributions and possible isolation of hubs
as a reflection of a local ``information horizon'' \cite{trusina2005}
with partial isolation between different biological processes. We here
address this problem by considering the yeast regulatory network
\cite{costanzo} with regards to protein properties. Using the Gene
Ontology (GO) Consortium annotations\cite{GO} we will show that
locality in the regulatory network primarily is associated to locality
in biological process, and only weakly related to functional abilities
of a protein.

\section*{Results}

Figure 1 show the regulatory network \cite{costanzo} for the yeast {\sl
  Saccharomyces Cerevisiae} and the color coded GO-graph for
annotations of biological processes. The GO-graph is colored such that
processes that are close are colored with similar colors.  The
proteins in the yeast network are then colored with the color of their
annotation, with hubs being colored according to the average of their
targets. If the targets of a given hub take part in a very broad range
of biological processes the color of the hub fades (gray). We see a
fairly scattered distribution of colors, with a tendency that proteins
in close proximity indeed are more similar.

More precisely, a GO-graph is an acyclic directed graph which organize
proteins according to a predefined categorization. A lower ranking
protein in a GO-graph share large scale properties with higher ranking
proteins, but are more specialized. In the GO-database, proteins are
categorized into three networks according to different annotations,
ranking known gene products after respectively: {\sl P)} biological
process, {\sl F)} functional ability/design of the protein and {\sl
  C)} cellular components where the protein is physically located. For
each of these {\sl three} ways of categorization we examined {\sl two}
distinct ways to measure GO annotation difference: the direct distance
and hierarchical distance (see box in Fig.\ 1).

Figure 2 presents the average GO-distance as function of distance $l$
in the regulatory network for each of the three different
GO-categories.  The regulatory distance is calculated by finding the
shortest path distance using breadth-first search disregarding the
directionality of the links. The
upper panels show that closely connected proteins are involved
in closely related cellular processes, {\sl P}. On the other hand,
the middle and lower panels show a weaker
relation between position in the regulatory network and
{\sl C} respectively {\sl F} based GO-distances.

In particular Fig.~2(a) shows that proteins separated by one or
two links are involved in similar processes. Here distance
$l=1$ mostly count proteins on the periphery of a hub and their
directly upstream and highly connected regulator. Distance $l=2$
count proteins regulated by the same highly connected
regulator. Note that we are averaging over all pairs in the whole
regulatory network including connections to less well-connected
regulators. In this way the highly connected nodes are counted for
each of their downstream targets and therefore the larger hubs will
make the dominant contributions to this calculation.

Figure 2(b) investigate the differences in GO-annotations,
but with the hierarchical distance that emphasize differences close to
the root of the GO-graph for processes(P). The fact that this measure
correlate to larger distances in the regulatory network implies that
proteins in a larger neighborhood of the regulatory network tends to be
on the same larger subbranches on the GO(P)-hierarchy.

In all the panels in Fig.\ 2 we also compare to a null model, generated
by keeping the regulatory network, but randomly reassigning which
proteins from the GO-graph that are assigned to which positions on the
network. This randomization maintain the positions of all nodes in the
regulatory network exactly. By doing this randomization one loose any
{\sl P, F} or {\sl C} correlation between a regulator and its
downstream targets. Any conceivable GO-distance therefore becomes
independent on the regulatory distance.

Figure 3 quantify the correlations observed for Fig.\ 2(a) and (b) by
comparing with another null model, which explicitly conserves the GO
annotations but allow for complete reorganizations of the
transcription network.  That is, we generate families of null models
by randomizing the regulatory networks while maintaining the in- and
out-degree for the nodes and with a bias for neighborhood correlations
of a GO annotation. In detail, for a bias parameter $\epsilon=0$, the
correlations are maximal given the available nodes in the original
network. For finite $\epsilon$ there are imperfections in the sampled
networks, which implies that there is some probability that the link
rewiring increases the GO distance.  Figure 3 show resulting
GO-distances as a function of distance in the yeast network for three
values of $\epsilon$.

From Fig.\ 3(a) we see that in order to reproduce the observed local
correlations of GO(P) in a random sample of networks, these need to be
generated with maximal bias. That is, the network generated with
$\epsilon=0$ reproduce observed correlations between processes of
proteins which are downstream of the same regulator {\it i.e.}\  at
distance $l=2$ in the regulatory network. At distances $l>2$
there are no detectable correlations, which in turn is reproduced by
allowing small imperfections ($\epsilon \sim 0.15$) in the rewiring.

In Fig.\ 3(b) we repeat the investigation from a), but with respect to
the hierarchical GO($P$) distance. In this case we see that $\epsilon
\sim 0.15 \rightarrow 0.30$ reproduce the observed correlations
between protein processes out to larger regulatory distances ($l\sim
3$). Figure 3(c)-(f), on the other hand, show that function or
cellular localization are only moderately related within the same hub
($l\sim2$), and unrelated at all larger distances.

\section*{Discussion}

Protein regulatory networks are highly functional information
processing systems, evolved to perform a diverse sets of tasks in a
close to optimal way. It is of no surprise that they are not
random, also in ways that can be detected without knowing much about
what actually goes on in the living system they regulate. However we
do not, a priori, know much about the relative importance of
function versus history:  
Is the topology of a network primarily governed by the processes it
direct, or is its topology influenced by random gene duplications
\cite{bhan,sole} and ``link" rewirings \cite{bornholdt}?

Concerning gene duplications
\cite{bhan,wagner,wu,maslov2004,teichmann,rodriguez,koonin,foster,enemark},
we detected 581 paralogous pairs among the 848 gene products in YPD,
see methods. Of these 581 pairs, only $\sim$15\% significantly
retained their common regulator, and only $\sim 0.6$\% of the proteins
pairs at distance $l=2$ are detectable pa\-ra\-logs. Therefore the
contribution from duplication events to any GO-similarity within hubs
can be ignored.

Our analysis in Figs. 2,3 emphasize the strong correlations between
network localization and process, in particular very strong (maximally
possible) correlation between process annotation of proteins in the
same hub. In addition, we see some functional similarities between
proteins in the same hub, in particular when considering the
hierarchical GO distances at $l=2$ in Fig. 3(d).  However we also find
that the functional diversity within hubs are large in terms of the
direct GO distance ($l=2$ in Fig. 3(c)).  Combined Fig. 3(c,d)
therefore show that proteins in the same hub have quite large direct
function-GO distances, but rarely belong to entirely different 
function-GO categories.


In any case we emphasize that we primarily find GO-processes
localized on hubs, and only weak correlations of
the functional abilities between proteins involved in the same
process.

The idea that process similarity are associated to network
localization is not new, and implicitly behind attempts to infer gene
networks from similarity in gene expression \cite{haverty}.  In the
supplement we use gene expression from micro-arrays to re-investigate
the correlation between process and locality in the regulatory
network. Thereby, we provide a broader support for our findings, and
present a quantitative illustration of the extent to which
gene-expression studies can be used to deduce co-regulation.

Support for the ubiquity of the ``one hub-one process" association is
also found from the fact that the likelihood that a regulatory protein
is essential is nearly independent on how many proteins it regulate
\cite{maslov2005}.  That is, the question of whether a null mutant of
a certain protein is viable is keyed to the essentiality of the
regulated process, and not to whether the process needs many or few
different ``tools" to be performed.

Overall we suggest that the topology of the yeast regulatory network
is governed by processes located on hubs, each consisting of a number
of tools in the form of proteins with quite different functional abilities.
This is consistent with a network evolution where gene duplication
occur, but where rewiring of regulatory links plays a bigger role
\cite{gu,berg,Ihmels,maslov2004,enemark}.  The regulatory network is
designed to co-regulate processes, and its evolutionary history of
must include a bias towards hub-regulation of individual processes.
Degree distributions are not broad because of duplication events, but
because a given biological task sometimes needs many, but typically
require few tools.

Finally our analysis have consequences for development of null models
for network topologies, and thereby for identifying functionally
important network motifs \cite{milo}.  While the previous null model
\cite{maslov2002} maintain in- and out- degrees of each protein, it
ignore correlations associated to cellular process.  When nearby
proteins are associated to the same processes one statistically expect
an increased probability for cliques \cite{Sneppen04a,bock}.  We
therefore expect that some of the many feed-forward loops in
transcription networks \cite{milo} will be explained by a new type of
null model: A null model where proteins contributing to a given
process are forced to remain close in the randomized network.

\section*{Methods}

The GO-annotations are used without any filtering. This does not
preclude bias introduced from using inferred annotations. Of the 848
genes in the YPD, 52 are not annotated and were thus not included in
the analysis. 142 genes has more than one molecular function, 314
genes takes part in more than one cellular component and 463 genes
participates in more than one biological process. To accommodate this
the analysis was carried out by choosing the annotations which
minimized the mutual distance for each pair of proteins. This choice
maximally resolves significant signals, since we minimize the effect
of the finite size of the GO-tree, and in the case of no signal this
choice introduces no bias.

Of the 848 gene products in YPD, we found 581 paralogous pairs using
BLASTP with E-value cutoff of $10^{-10}$~\cite{maslov2004,bock07}. For
the YPD network 132 of these paralogous pairs are at distance
$l=2$. This should be compared to a null expectation of $50\pm 6$
paralogous pairs at $l=2$ found by randomizing the YPD network while
keeping in- and out-degrees \cite{maslov2002}. Therefore at max
132-50=82 of the paralogous pairs are in the same hub due to their
history of common origin.  This correspond to $82/581 \sim$15\% of
duplicated proteins in YPD.  The excess of 82 paralogous pairs at
distance 2 should also be compared to the total of 13554 protein pairs
that the YPD network have at distance $l=2$. Thus only $\sim 0.6$\% of
all proteins pairs at $l=2$ are detectable paralogs.

As seen in our supplementary material, we reach the same basic
conclusion of hubs being functionally isolated using a completely
different approach based on gene expression data. Ana\-ly\-zing
micro-array data from 482 stress experiments from Saccharomyces Genome
Database (\verb+www.yeastgenome.org+) using thresholds methods from
\cite{benjamini95} we indeed find localization of perturbations on our
regulatory network.  Thus the appendix support the robustness of our
results to an independent categorization of protein processes.

\section*{Acknowledgements}
We acknowledge the support from the Danish National Research
Foundation through ``Center for Models of Life'' at the Niels Bohr
Institute. KS and JBA wishes to thank the Lundbeck Foundation for
funding PhD-studies.  JBA wish to thank The Fraenkel Foundation.




\clearpage

\noindent {\bf Fig. 2.} GO-distance between two nodes as a function of
separation in number of steps in the regulatory network of {\sl
S.Cerevisiae} \protect{\cite{costanzo}}. The upper middle and lower
panel refer to respectively the {\sl Process, Function} and
{\sl Component} GO-annotation. In left and right side of figure we
analyze respectively the direct GO-distance and the hierarchical
GO-distance, explained in box under Fig. 1.

\clearpage

\begin{figure}[!htp]
\includegraphics[width=.45\columnwidth]{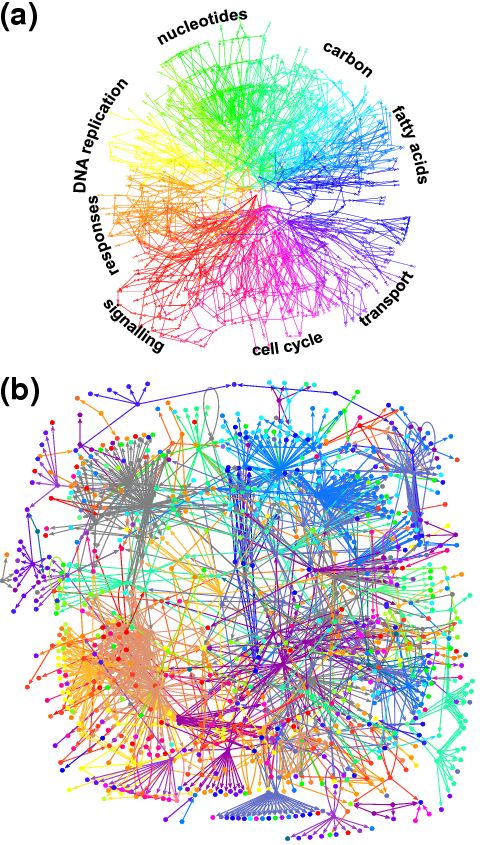}
\includegraphics[width=.50\columnwidth ]{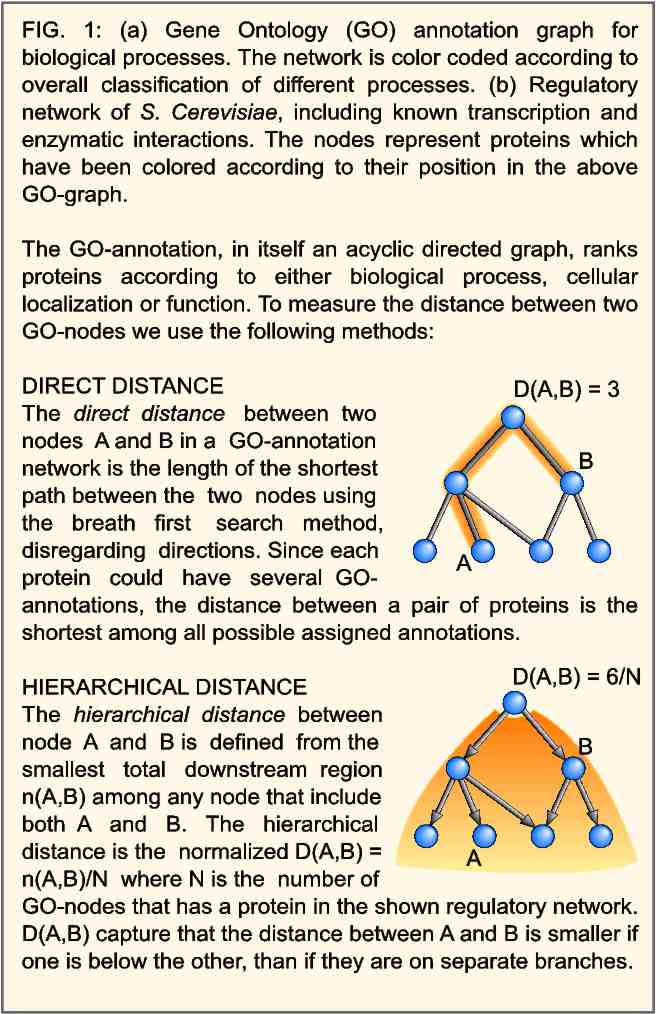}
\label{fig1}
\end{figure}

\begin{figure}[!htp]
{\large Figure 2}
\begin{center}
\includegraphics[width=.75\columnwidth ]{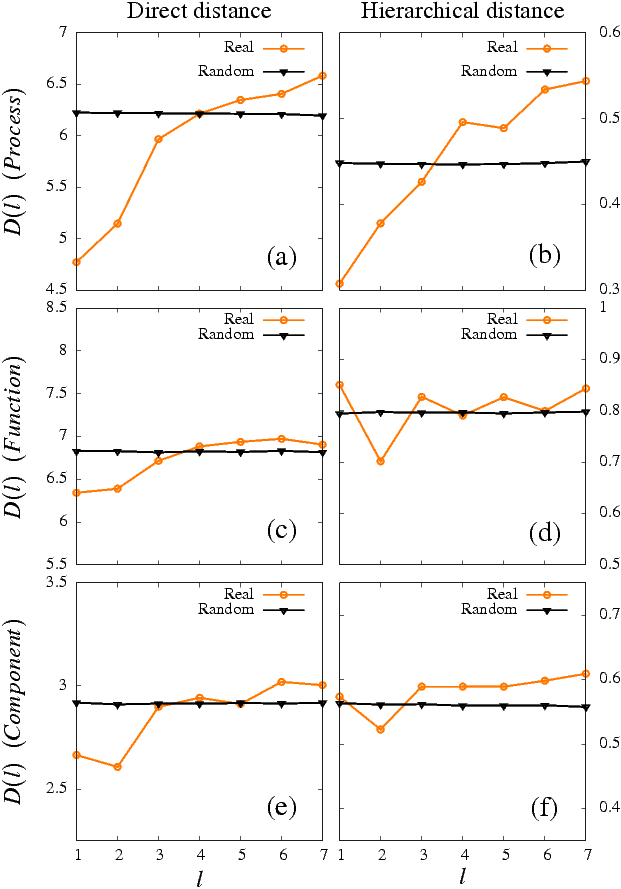}
\end{center}
 \label{fig2}
\end{figure}

\begin{figure}[!htp]
\centering
\includegraphics[width=.5\columnwidth ]{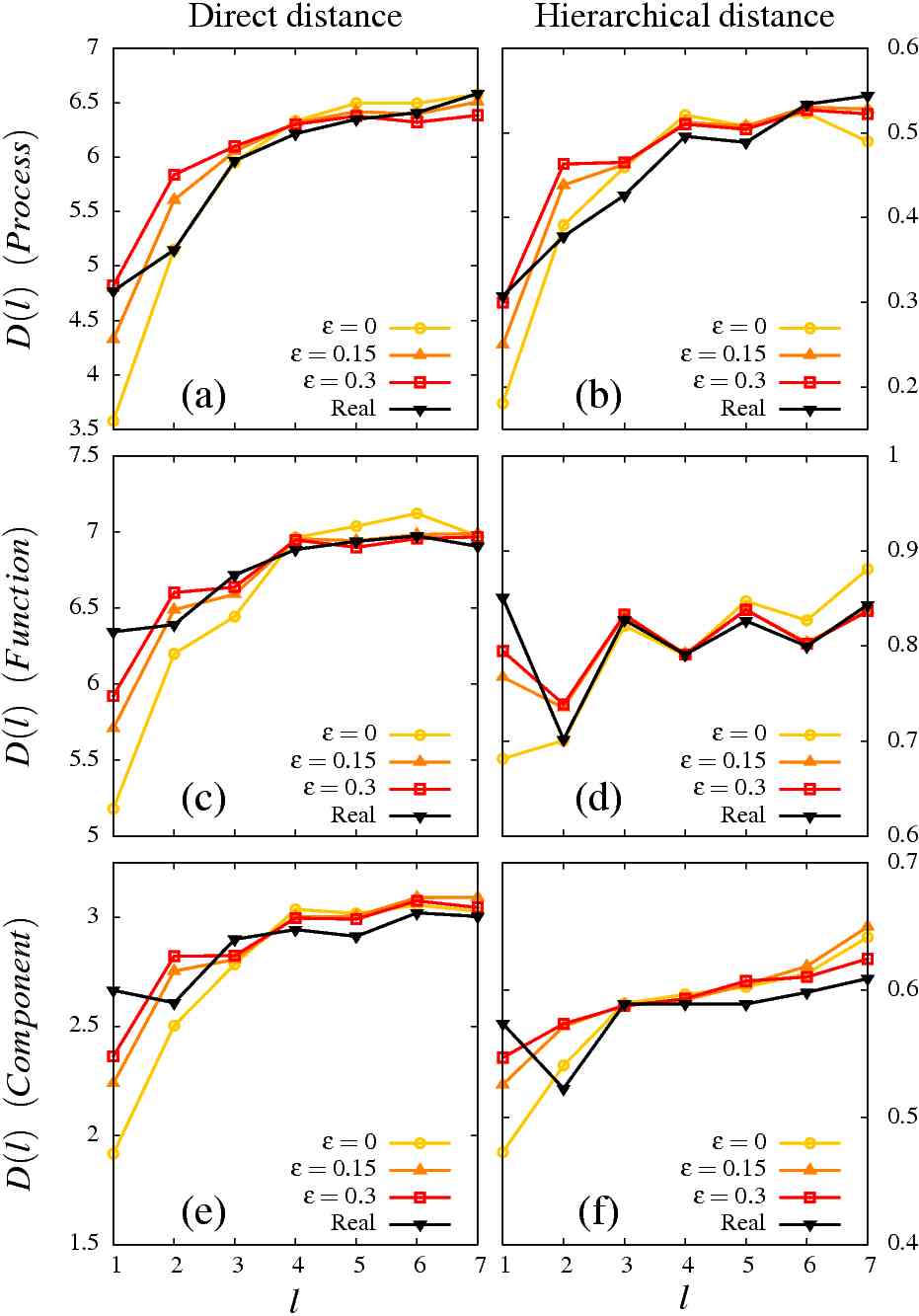}\\
\includegraphics[width=.5\columnwidth ]{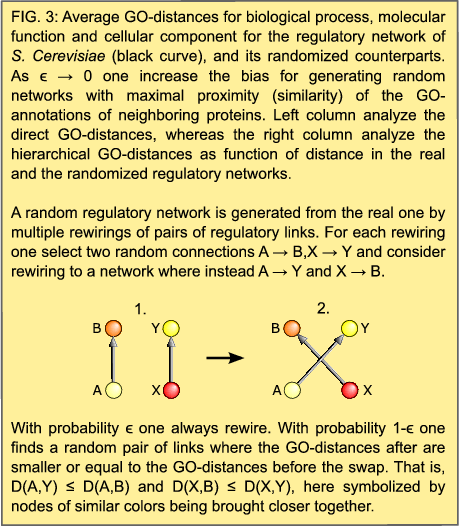}
\label{fig3}
\end{figure}

\clearpage
\noindent
\section{Supplementary material}

\begin{figure}[!htp]
\begin{center}
\includegraphics[width=.3\textwidth]{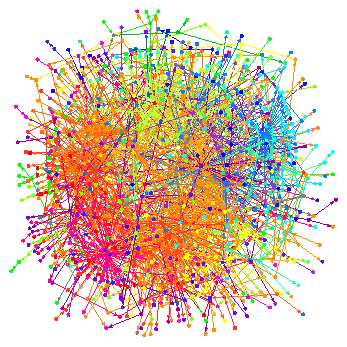}
\includegraphics[width=.3\textwidth]{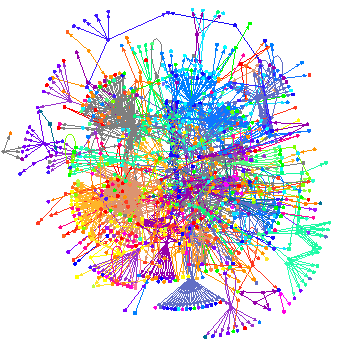}
\includegraphics[width=.3\textwidth]{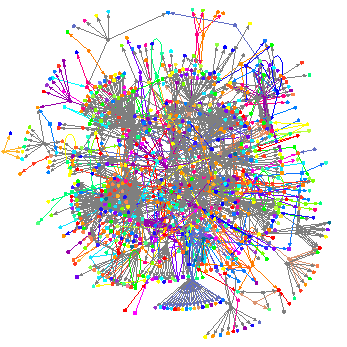}
\end{center}

\caption{ (a) Rewiring the YPD network by bringing nodes of similar
  biological process annotation closer together. (b) The real YPD
  network. (c) The YPD network with randomized GO annotation for
  biological processes.}
\label{fig:YPD_vers}
\end{figure}

\noindent
In Fig.~\ref{fig:YPD_vers} we have shown alternative ways of
communicating the findings in the paper. In Fig.~\ref{fig:YPD_vers}(b)
we display the YPD network from Fig.1 in the main paper. To visualize
that the GO annotations are indeed optimized and non-random we have in
Fig.~\ref{fig:YPD_vers}(c) shown the YPD network where we have
reshuffled the GO annotations. Since the hubs are annotated to a
collection of randomly selected biological processes they will all
appear as overall ambiguous (gray) while all enzymes (end nodes)
appears perfectly mixed in a harlequin-like fashion. In contrast, the
clearly separated functional neighborhoods in the real network in
Fig.~\ref{fig:YPD_vers}(b) leaves most of the hubs with a easily
identifiable color. Conversely, it is natural to ask if the YPD
network is optimized according to GO-annotation proximity. In
Fig.~\ref{fig:YPD_vers} we have rewired the YPD network in order to
bring nodes that share biological processes closer together. Compared
to the real network the effect is a clearer functional separation, but
there is a trade-off in the resemblance of the topology.

Ref.~\cite{bock} and~\cite{Sneppen04a} investigated the hierarchical
properties of networks with broad-degree distributions.  The hierarchy
index ${\cal F}$ was defined as the fraction of hierarchical paths out
of the total number of paths. A hierarchical path is a path where the
hierarchy is preserved along the path in the sense that low-ranking
nodes always receive orders and never gives orders.

Using the hierarchy index, ${\cal F}$, we can clarify the trade-off in
Fig.~\ref{fig:YPD_vers}(a) by calculating the ${\cal F}$ value
disregarding directionality and finding it to be ${\cal F}=.49(3)$.
The real YPD network in Fig.~\ref{fig:YPD_vers}(b) has ${\cal F}=.26$.
The network in Fig.~\ref{fig:YPD_vers}(c) is per definition
topologically identical to the real network in
Fig.~\ref{fig:YPD_vers}(b). For an ensemble of randomly rewired
versions of the YPD network we found ${\cal F}=.60(3)$. Thus, when
optimizing the YPD network according to biological process
neighborhoods we partly loose the observed hub-hub separation.
Therefore we conclude that the model is extremely simple and is not
able to capture both functional neighborhood optimization and soft
modularity.

We also investigated the relationship between locality and function by
the use of microarray results from yeast cultures subjected to
different stress conditions. This choice was motivated by the fact
that expression of enzymes during stress shows the strongest response
in terms of variance and expression fold. In contrast, the variance
and fold change of regulators regulating other regulators has low
signal-to-noise using microarrays.

482 microarray stress experiments on {\it Saccharomyces Cerevisiae}
were downloaded from Stanford Genome Database
(\verb+www.yeastgenome.org+). These experiments covered the following
conditions: heat shock, osmotic shock, diauxic shift, hydrogen
peroxide shock, Menadione, diamide, sorbitol, DTT, amino acid
starvation, MMS, Nitrogen depletion, Na+, Phosphate, Sulfate, Uracil,
sugar enriched media, gamma irradiation, YPD medium inoculation and
others. The data was centered for each gene.

We hypothesized that the correlation between expression of a given
protein and its network neighborhood would show the strongest average
signal if we focused on the hubs and their targets.
That is, the large number of target proteins of hub-regulators will
enhance any functional locality in the signal compared to the bulk
signal at larger distances.

We compared the overlap of all pairs of hubs larger than size 10 and
then removed the hubs that had a larger target overlap than 1/3 of the
targets of another, larger hub. This ensured that functional response
overlap would not be from network overlap. In this way we ended up
with a total list of 26 isolated hub regulators from the regulatory
network of yeast.

\begin{figure}[!tp]
\includegraphics[width=.50\textwidth]{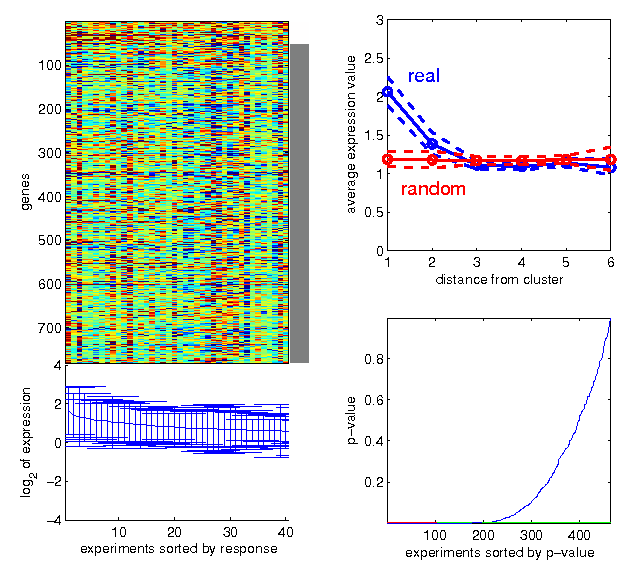}
\includegraphics[width=.45\textwidth]{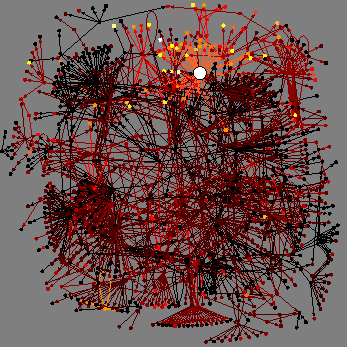}

\caption{The responsiveness filter (here shown for the weakest responding
  cluster): sort experiments according to response in selected cluster
  and perform {\sl t-}test with 0.5\% p-value cut-off, correct for
  false discoveries and finally only accept two-fold responses. Left:
  the data for the cluster (top of matrix) and the rest of the genes
  (bottom/gray sidebar). The color scheme is red for expression
  two-fold enhanced, green for zero change and blue for two-fold
  depressed. The curve and error bars are the average experimental
  values in the cluster sorted in descending order from right to left
  (in $\log_2$ transformed format). Middle: the average response
  resolved by distance from the hub in question, here the hub is GCN4.
  The blue curve and dashed errors is the real signal, and the red
  curve is a random expectation created by randomly swapping the
  expression data for the genes.  There is a clear signal for the
  targets of the hub and a weak signal for two-steps away.  Mid-low is
  the false discovery management procedure as referred to in the text.
  Right: the response mapped onto the Yeast Regulatory network with a
  ``hot" color scheme where light yellow is strong response and dark
  red is no response. The tested cluster shows a clear locality
  resolved response.}

\label{fig:FDR}
\end{figure}

We detected responsiveness of a hub by a combination of statistical
tests and biological reasoning. The target set of enzymes of a hub was
considered as a supervised cluster to be tested against the bulk of
the network. The variance of each experiment in the cluster versus the
rest of the network was analyzed by using a {\sl t-}test. To manage
the false discovery rate for the multiple comparisons we use the
method of ~\cite{benjamini95} with a conservative p-value threshold of
0.5\%. If any of the resulting, significantly upregulated experiments
showed at least a two-fold upregulated response as well, the hub was
dubbed responsive to stress. In this fashion only 12 out of the 26
selected hubs were found responsive. Figure \ref{fig:FDR} shows the
total average response resolved by distance. In this figure the
cluster is responsible for amino acid synthesis according to the
available biological process GO-annotation. Our analysis finds that
the experiments that most strongly activates this cluster are amino
acid starvation, and nitrogen depletion type experiments. This
supports the overall postulation of the paper that network locality
means functional locality and thus points to local hierarchies of hubs
and their targets as natural ``soft'' modules in biological
regulation.

\begin{figure}[!t]
\begin{center}
\includegraphics[width=.8\textwidth]{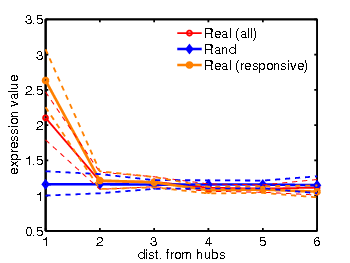}
\end{center}
\caption{
  Total average response to stress in the yeast regulatory network.
  The blue diamonds is the random expectation created by random
  swapping the expression data for the genes. The red circles is the
  average response for all 26 clusters for the 20 first experiments in
  the sorted list. The orange dots is the average response for the 12
  clusters that were found to be responsive according to our
  criterion(see text). As can be seen the signal is clear for local
  clusters and then fades rapidly further away.}

\label{fig:Tot}
\end{figure}

For the responsive hubs the experiments were sorted in descending
order according to the average upregulation in the cluster. For the 20
first experiments in each list the average response per distance was
calculated. Notice that we here include experiments irrespectively of
the outcome of the responsiveness filter. This is to avoid selecting
experiments that {\sl only} activates the cluster and nothing else, a
reasonable choice since we are interested in functional locality of a
neighborhood at different distances. In Fig.~\ref{fig:Tot} we show the
total average response of the stress responsive hubs resolved by
distance and compared to random expectation.  In this figure we see a
clear local functionality on average, which naturally does not account
for co-activation effects.

\begin{figure}
\begin{center}
\includegraphics[width=.8\textwidth]{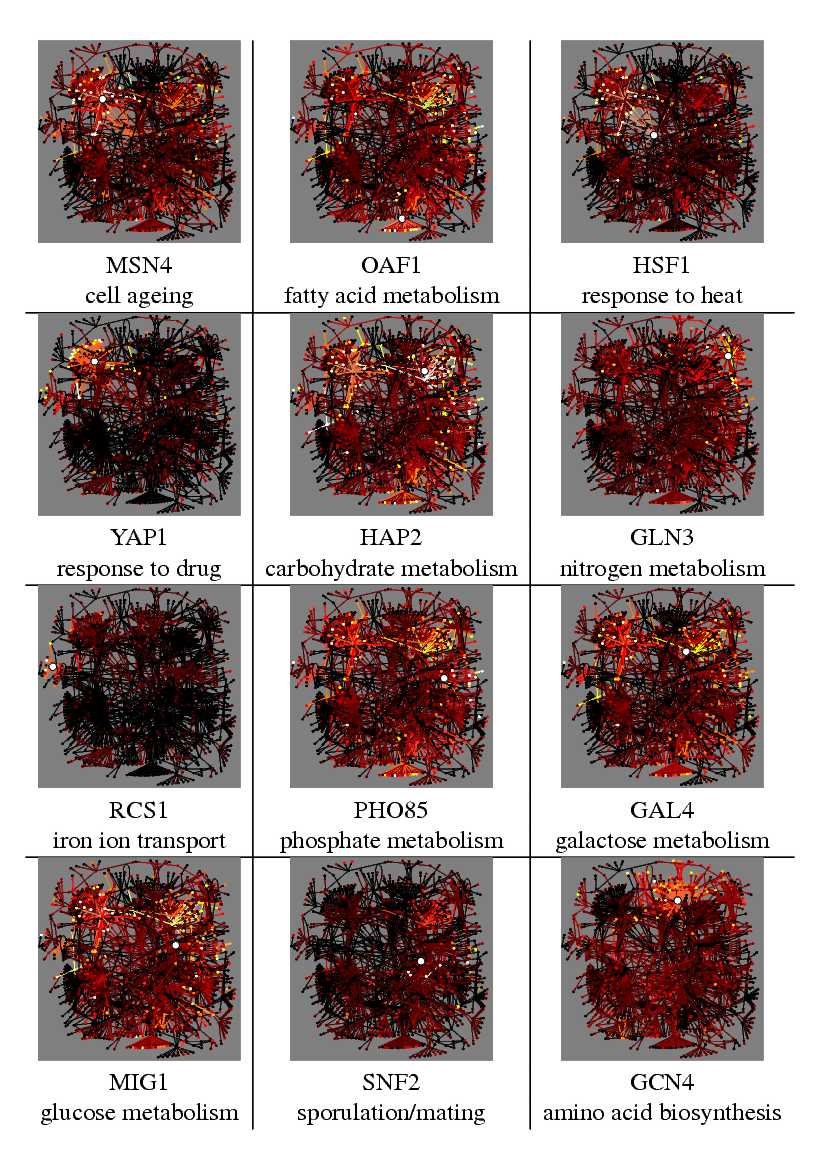}
\end{center}
\caption{
  The 12 most responsive hubs and clusters. The organization is:
  strongest response in the upper left corner and then descending to
  lower right corner. The name of the hub and the biological process
  annotation is indicated below each graph. The color scheme is "hot"
  meaning strongest response is white and weakest response is dark
  red. The hub of the investigated cluster is enhanced and colored
  white for each graph. The locality of the signal is clear, but there
  is also often a clear co-activation.  Further, as can be seen there
  is a region, located around the middle and lower left of the network
  which is almost always silent in stress conditions. }
\label{fig:all}
\end{figure}

\begin{figure}
\begin{center}
\includegraphics[width=.65\textwidth]{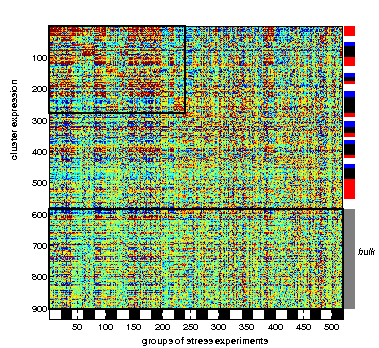}\\
\includegraphics[width=.65\textwidth]{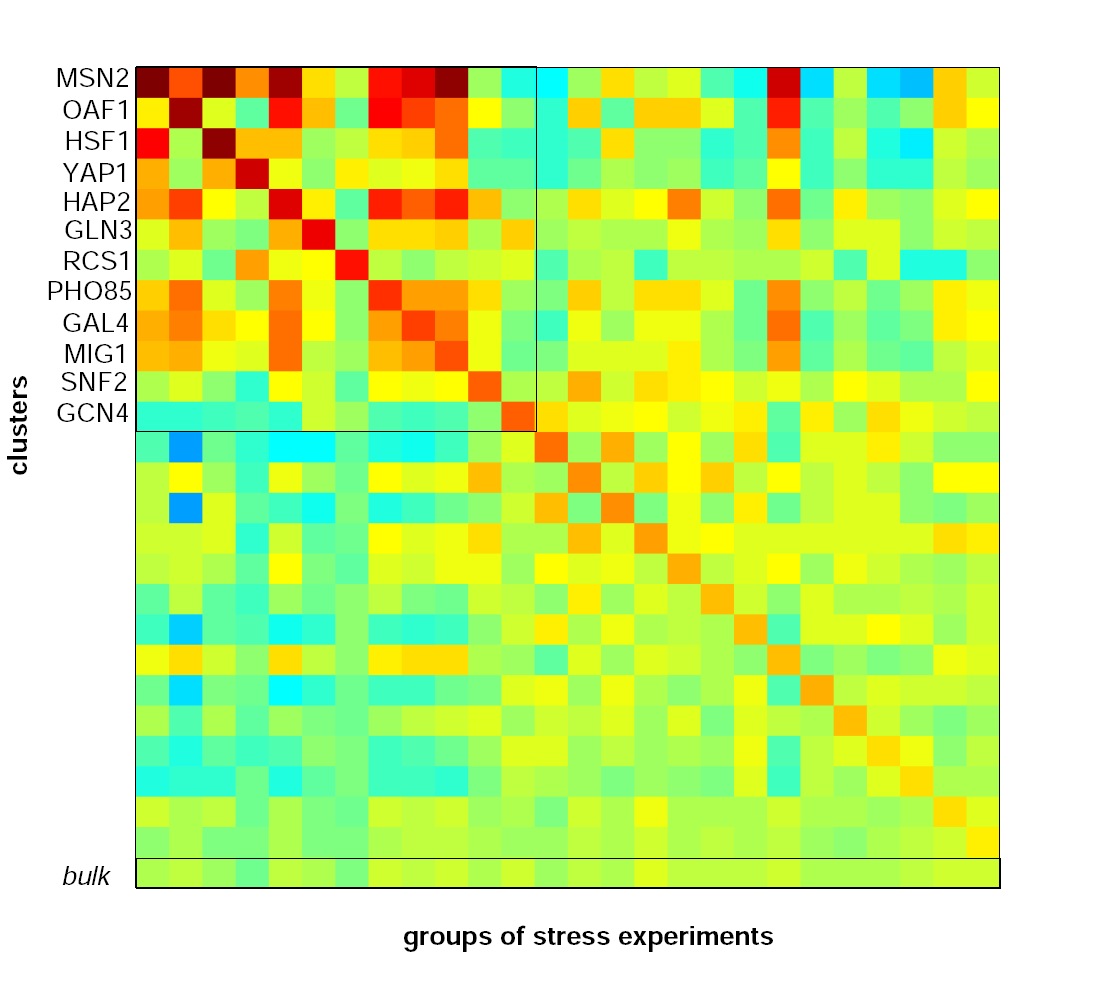}
\end{center}
\caption{
  Investigating co-activation of clusters sorted by response strength
  descending from upper left corner. Top: the raw data for each of the
  26 clusters. The color bar to the right delineates the groups of
  genes belonging to each cluster. The black and white color bar at the
  bottom demarcates the groups of the 20 strongest activating
  experiments for each cluster. Bottom: the raw data has been
  coarse grained by taking the average in each bin. The name of the
  cluster has been indicated to the right. For both matrices the
  responsive clusters are indicated with the black box at the upper
  left corner. The black box at the bottom is the response of the
  genes not in any cluster. The color scheme is red for strong
  activation (larger than 2), blue for deactivation (less than .5) and
  green for no activation. }
\label{fig:co}
\end{figure}

In Fig.~\ref{fig:all} we show the stress response mapped onto the
network for each cluster. The co-activation of clusters is obvious.
The annotation of the biological processes also raises the expectation
of such co-activation. For example, HAP2, GAL4 and MIG1 are all
related to the utilization of carbon sources and would be expected to
co-activate to a certain degree. This is somewhat visible in the
graphs as an overall activation of the same regions for those three
hubs.

A more tangible analysis of the modular co-activation is shown in
Fig.~\ref{fig:co}. Here the raw data is shown in an expression matrix
with rows indicating genes and columns indicating experimental
conditions. The second matrix is a coar\-se-grain\-ed version of the
raw data for ease of analysis. Next to the raw data matrix is a
color bar indicating the number of genes in each cluster including the
non-responsive ones. A black box in the upper left corner separates
the responsive clusters from the rest. Focusing on the coarse grained
lower matrix the picture becomes clear: there is a large group of of
more or less co-activated clusters with MSN2, OAF1, HSF1, HAP2, PHO85,
GAL4 and MIG1 as central members, YAP1 and GLN3 as semi-correlated
members and three clearly independent clusters: RCS1, SNF2 and GCN4.
Since we have removed network overlap in the initial choice of hubs
and clusters the co-activation stems from a combined reaction to the
most stressful conditions. As expected above the clusters HAP2, GAL4
and MIG1 are indeed seen to be co-activated in this analysis. The
co-activation clusters are often separated in the network, thus
underpinning the point about the modules serving as tools for the
organism to be employed according to need.

Finally we investigating the full biological process GO-annotation for
the 12 clusters along with the SGD experiment access codes and
description.
It becomes clear that the large co-activated cluster arises from heat
shock experiments that triggers protein folding responses (HSF1), cell
aging (MSN2), membrane reconstitution (OAF1) and energy production
(HAP2, PHO85, GAL4 and MIG1). This is a coordinated response resulting
from external stimuli triggering many independently regulated needs.
Furthermore, the semi-independence of YAP1 is the specialized tools
that the regulator controls and thus it is mostly triggered by
hydrogen peroxide and diamide shocks. The semi-independence of GLN3
comes mostly by starvation of nitrogen, amino acids and adenine. The
purely independent response of RCS is ambiguous, since none of the
stress conditions in our database matches the iron ion function of
this cluster. SNF2 is seemingly activated by phosphate depletion which
triggers a mating-type phenotypic switch. And finally, GCN4 is clearly
activated by a pure setup of amino acid and nitrogen starvation
conditions creating a demand for biosynthesis of these compounds.


%
%
%
%
%
%
%
%
%
%
%
%
%
%
%
%
%
%
\end{document}